\documentstyle[psfig]{mn}

\newif\ifAMStwofonts

\newcommand{\msun}{M_{\odot}}

\title[]{The standstill luminosity in Z Cam systems}
\author[]{R.~Stehle, A.~King, C. Rudge \\
           Astronomy Group, University of Leicester, Leicester,
               LE1 7RH, UK}
\date{Accepted. Received}
\pagerange{\pageref{firstpage}--\pageref{lastpage}}
\pubyear{2000}

\begin{document}

\label{firstpage}

\maketitle

%------------------------------------------------------------
%============================================================
%------------------------------------------------------------

\begin{abstract}
We consider accretion discs in close binary systems.
We show that heating of a disc at the impact point of 
the accretion stream contributes significantly to the local energy
budget at its outer edge. As a result the thermal balance relation
between local accretion rate and surface density (the `S--curve') changes;
the critical mass transfer rate above which no dwarf
nova outbursts occur can be up to 40\% smaller than
without impact heating. Standstills in Z~Cam systems thus
occur at smaller mass transfer rates than otherwise expected, 
and are rather fainter than the peak luminosity during the dwarf
nova phase as a result.
\end{abstract}


\section{Introduction}

\label{visc_intro}

Dwarf novae are a subset of cataclysmic variables (CVs), close binary
systems in which a white dwarf accretes from a low--mass
secondary (see Warner, 1995). 
Their defining feature is outbursts lasting a few days,
which recur at intervals of weeks.  The thermal--viscous disc
instability model is generally accepted as the most successful
explanation for these outbursts (see Cannizzo 1993 or Warner 1995 for
recent reviews).  The outbursts are thought to result from a thermal
instability of the disc when hydrogen is partially
ionized. Hence accretion cannot be steady for a certain range of mass
transfer rates $-\dot M_2$ from the secondary such that dissipation at
local accretion rates $\dot{M}_{\mathrm{acc}} = -\dot M_2$ implies
partial ionization somewhere in the disc. We can write this condition
as $ \dot{M}_{\mathrm{B}} \le - \dot{M}_{2} \le \dot{M}_{\mathrm{A}}$,
where $\dot{M}_{\mathrm{A}}$ and $\dot{M}_{\mathrm{B}}$ are the
critical local accretion rates for low and high ionization at a given
point of the disc (see Fig. 1). 

If this condition holds, 
the disc goes through a limit
cycle with alternating high ($\dot{M}_{\mathrm{out}}$) 
and low ($\dot{M}_{\mathrm{qui}}$) mass accretion rates.
If the mass transfer rate is higher than 
$\dot{M}_{\mathrm{A}}(r)$ everywhere in the disc, 
the accretion disc is in permanent outburst and 
$\dot{M}_{\mathrm{acc}}= - \dot{M}_{2}$ is stable.
As the critical mass accretion rates increase with radius
this condition is equivalent in saying that accretion discs are stable
and 
% begin 8.Sep 99 
the whole disc is
% end   8.Sep 99 
permanently in outburst as long as 
\begin{equation}
-\dot{M}_{2} > \dot{M}_{\mathrm{A}}(r_{\mathrm{disc}})
\label{1}
\end{equation}
where $r_{\mathrm{disc}}$ is the radius of the outer disc edge. (A
second stable regime exists if 
$- \dot{M}_{2} < \dot{M}_{\mathrm{B}}(r_{\mathrm{in}})$, where 
$r_{\mathrm{in}}$ is the radius of the inner edge of the accretion disc.
This in practice, however, requires unrealistically low mass transfer
rates.)

To lowest order $\dot{M}_{\mathrm{out}}$ and $\dot{M}_{\mathrm{qui}}$
are independent of $ - \dot{M}_{2}$. Thus one can approximately divide
systems into steady (novalike) systems or dwarf novae according as
(\ref{1}) holds or not, with $\dot{M}_{\mathrm{A}}(r_{\mathrm{disc}})$
a universal function of outer disc radius.

\begin{figure}
  \centerline{\psfig{figure=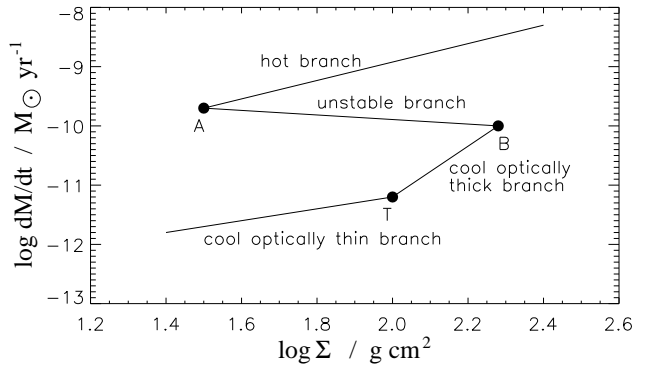,width=9cm}}   
  \caption{A typical $f$--$\Sigma$ relation at
 a disc radius $r \sim 10^{10}$cm.}
\end{figure}

Z~Cam systems are a subset of the dwarf novae which occasionally show
standstills, i.e. epochs where the luminosity is constant for several
days, at a value intermediate between the outburst and quiescent
states 
(e.g. Warner, 1995, 
% begin 8.Sep 99 
Oppenheimer, Kenyon \& Mattei 1998, 
Section 3.4),
and a mean brightness during standstills exceeding the time
averaged brightness during the dwarf nova phase
(Honeycutt et al. 1998).
% end   8.Sep 99 
One can qualitatively understand these as systems where the
mass transfer rate $-\dot M_2$ is close to the critical value
$\dot{M}_{\mathrm{A}}(r_{\mathrm{disc}})$, but is somewhat
time--variable. Thus when $-\dot{M}_{2} >
\dot{M}_{\mathrm{A}}(r_{\mathrm{disc}})$ the accretion disc is
permanently in outburst, corresponding to a standstill, while at times
when $-\dot{M}_{2} < \dot{M}_{\mathrm{A}}(r_{\mathrm{disc}})$ the
system undergoes a limit cycle as a normal dwarf nova system. Cannizzo
\& King (1998) suggested this type of picture, with the required
variations in $-\dot M_2$ resulting from a varying population of
starspots near the $L_1$ point on the secondary.

Although this picture is attractive, Esin et al. (2000) point out a
quantitative shortcoming: it appears to predict standstill
luminosities which are as bright as (or brighter than) the peak
luminosity during dwarf outbursts. For an outbursting disc behaves as
if steady, i.e. the accretion rate is almost constant through it. At
the peak of the outburst, just before the cooling wave propagates
inwards, the accretion rate at the outer edge of the hot disc region
(which may be the outer edge of the whole disc) is by definition very
close to the value $\dot{M}_{\mathrm{A}}(r_{\mathrm{disc}})$. Hence we
expect that $\dot{M}_{\mathrm{out}} \simeq
\dot{M}_{\mathrm{A}}(r_{\mathrm{disc}})$ However, standstills occur
only when $\dot{M}_{\mathrm{2,standstill}} \ga
\dot{M}_{\mathrm{A}}(r_{\mathrm{disc}})$. Thus one might expect that
\begin{equation} 
\dot{M}_{\mathrm{standstill}} \ga \dot{M}_{\mathrm{out}}
\label{2}
\end{equation}
resulting in a comparable or greater disc brightness during
standstills. This is in direct contradiction to observations which
always show that the disc brightness during standstills is lower than
the peak of any outburst.

It is clear that the derivation of the discordant inequality (\ref{2})
relies on the assumption that the critical mass transfer rate
$\dot{M}_{\mathrm{A}}(r_{\mathrm{disc}})$ is essentially independent
of $-\dot M_2$. We show here that this is not correct, since the disc
is heated at a local rate $Q^+_{\rm HS}$
by the impact of the gas stream from the secondary, and this
heating effect increases as $-\dot M_2$ increases. 
% begin 8.Sep 99 
$\dot{M}_{\mathrm{A}}$ therefore depends not only on the 
the disc radius, but also on the mass transfer rate, i.e.
\begin{equation}
\dot{M}_{\mathrm{A}}=
\dot{M}_{\mathrm{A}}(r_{\mathrm{disc}}, \dot{M}_{2}).
\end{equation}
The critical mass transfer rate is then given by the solution 
of the implicit equation
\begin{equation}
\dot{M}_{2} = \dot{M}_{\mathrm{A}}(r_{\mathrm{disc}}, \dot{M}_{2}).
\end{equation}

We show explicitly in the next section that 
$\partial \dot{M}_{\mathrm{A}}(r_{\mathrm{disc}}) / \partial Q^+_{\rm
HS} < 0 $.
Taking this extra  heating by the hot spot into account,
the condition for the accretion disc to be permanently in outburst 
is already fulfilled at much smaller values than indicated by 
the unperturbed value of $\dot{M}_{\mathrm{A}}$.
Outbursts for mass transfer rates smaller than 
the hot spot heating condition are not dramatically 
altered by the new outer boundary condition, so
we conclude that $\dot{M}_{\mathrm{standstill}}$ 
is considerably lower than 
$\dot{M}_{\mathrm{out}}$, and hence that standstills should be
fainter than outbursts.

%=======================================================================
%------------------------------------------------------------

\section{The $f$-$\Sigma$ relation with impact heating}

\begin{figure*}
  \centerline{\psfig{figure=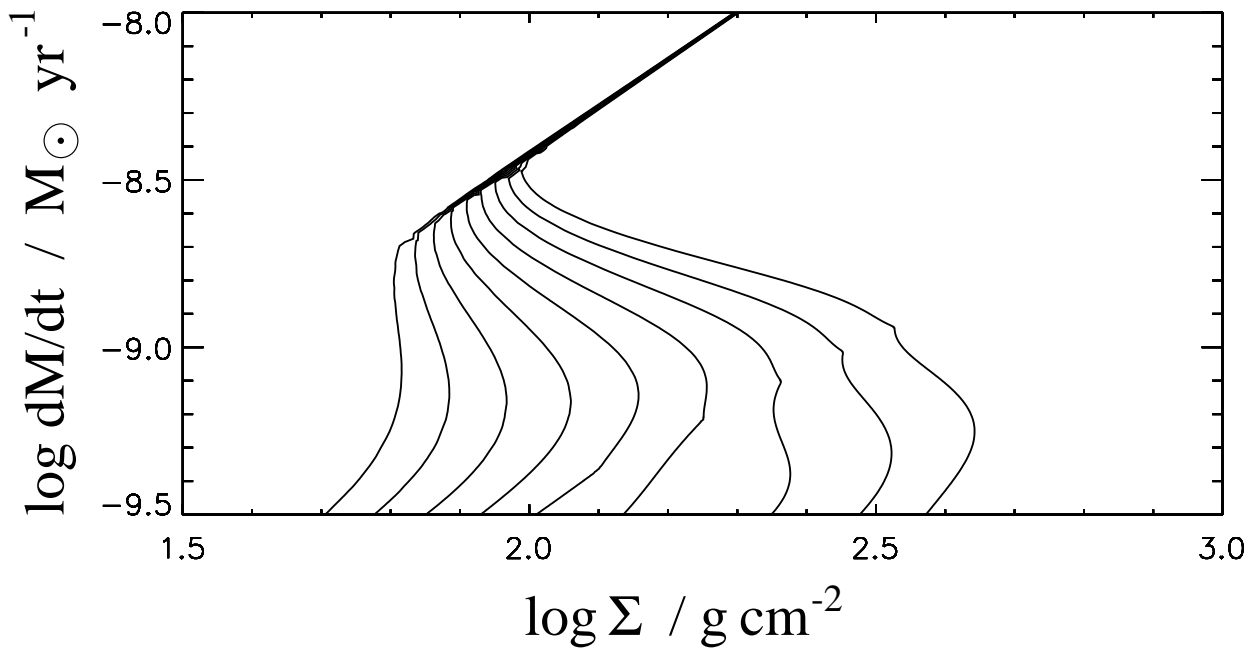}}   
  \caption{The $f$--$\Sigma$ relation for $r_{\mathrm{disc}} =2.7 \,
  10^{10}$cm and impact heating of $Q^{+}_{\mathrm{HS}}/10^{10}$
  erg s$^{-1}$ cm$^{-2} = (0, 1, 2,
  4, 6, 8, 10, 12, 14, 16)$ from right to left.}
\end{figure*}

The so called $f$--$\Sigma$ relation for a disc is defined as the locus
in the $\dot{M}_{\mathrm{acc}}$--$\Sigma$ plane
where the local heating by small--scale viscous processes, i.e.
\begin{equation}
Q^{+}_{\mathrm{visc}} = \frac{9}{4} 
\frac{G M_{\mathrm{WD}}}{r_{\mathrm{WD}}^{3}} f
\end{equation}
is in equilibrium with radiative cooling of both surfaces
of the accretion disc, given by 
\begin{equation}
Q^{-}_{\mathrm{rad}} = 2 \sigma T_{\mathrm{eff}}^{4}.
\end{equation}
Thus we have
$\dot{M}_{\mathrm{acc}}=3 \pi f$, with
$f = 2/3 \alpha c_{\mathrm{s}} H_{\mathrm{eq}} \Sigma$
(cf. Ludwig \& Meyer 1998, among others).

The thermal equilibrium curve $Q^{+}_{\mathrm{visc}} =
Q^{-}_{\mathrm{rad}}$
is determined by the vertical temperature structure in 
the accretion disc through the requirement
$T_{\mathrm{eff}} = T_{\mathrm{eff}} (T_{\mathrm{c}})$ at 
a given surface density $\Sigma$.
We follow Ludwig \& Meyer (1998) and assume that 
the relation between $T_{\mathrm{eff}}$ and 
$T_{\mathrm{c}}$ is independent of the actual heating 
mechanism.

\begin{figure}
  \centerline{\psfig{figure=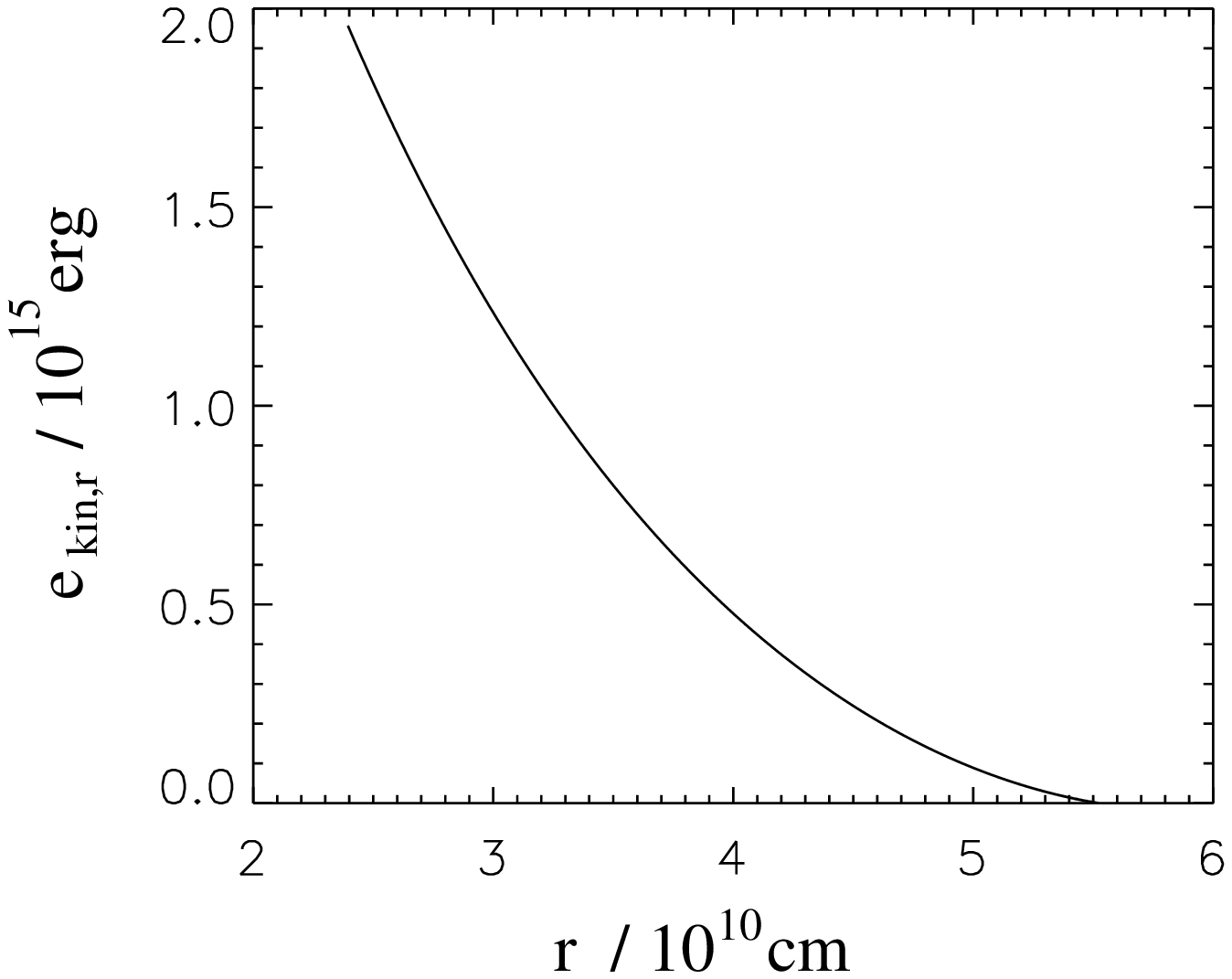,width=9cm}}   
  \caption{The radial kinetic energy $e_{\mathrm{kin,r}}$ of a ballistic
particle for 
           $M_{1}=1.15 \msun$, $M_{2}=0.67 \msun$ and
           $P_{\mathrm{orb}}=3.8$h.}
\end{figure}

In Fig. 2 we show the $f$--$\Sigma$ relation for the parts of the
accretion disc where the energy impact from the ballistically 
falling accretion stream particles contribute significantly to the local
energy budget, i.e. where 
\begin{equation}
Q^{+}_{\mathrm{visc}} + Q^{+}_{\mathrm{HS}} = Q^{-}_{\mathrm{rad}}.
\end{equation}
Here the impact heating is given by 
\begin{equation}
 Q^{+}_{\mathrm{HS}} = - \eta_{\mathrm{heat}}
\frac{ e_{\mathrm{kin,r}} \dot{M}_{2} }{2 \pi r \delta r} = \mbox{const.}
\end{equation}
where $e_{\mathrm{kin,r}}$ is the kinetic energy of the ballistically
infalling mass thermalized at the impact of the accretion stream. We
assume that only the radial velocity component of the transferred
material is thermalized, with efficiency $\eta_{\mathrm{heat}}$; for a
strong impact shock $\eta_{\mathrm{heat}}$ is close to unity.  Thus
$Q^{+}_{\mathrm{HS}}$ is of order a few times $10^{15}$erg g$^{-1}$
for typical binary and disc parameters (see Fig. 3).  The impact point
is stationary in the corotating binary frame, and heats a ring $2
\pi r_{\mathrm{disc}} \delta r$ of disc material at the outer edge,
where $\delta r$ is the radial extent of the impact point, typically
of the order of a few percent of $r$.

For typical binary and disc parameters we find
\begin{equation}
Q^{+}_{\mathrm{HS}} \simeq 1.5\times
10^{10}\eta_{\mathrm{heat}}\dot{M}_{16}
 \mbox{erg s}^{-1} \mbox{cm}^{-2},
\label{5}
\end{equation}
where $\dot M_{16} = -\dot M_2/10^{16}$g s$^{-1}$. Fig. 2
shows $\dot{M}_{\mathrm{acc}}$ as a function of 
$Q^{+}_{\mathrm{HS}}$ at a disc radius of $2.7 \times 10^{10}$cm 
for a 1.15$\msun$ white dwarf.
Using the parameterization (\ref{5}) of 
$Q^{+}_{\mathrm{HS}}$
we derive a critical mass transfer rate from 
\begin{equation}
\log \left(
\frac{\dot{M}_{\mathrm{A}}} {M_{\sun} \mbox{yr}^{-1}} \right) \simeq
-8.48 - 0.013 \frac{Q^{+}_{\mathrm{HS}}}{10^{10} \mbox{erg}\,
\mbox{s}^{-1} \, \mbox{cm}^{-2}} 
\end{equation}
for $Q^{+}_{\mathrm{HS}}<17 \times 10^{10}$erg cm$^{-2}$ s$^{-1}$.
The S--curve is straightened out for impact energies greater 
than $17 \times 10^{10}$erg cm$^{-2}$ s$^{-1}$ and no thermal--viscous
instability occurs close to the outer disc edge.

We see that including heating by the accretion stream lowers the
threshold mass transfer rate preventing a return to quiescence by
about 40\%.
%lower 
%for $\eta_{\mathrm{heat}}=1.0$, 
%28\% for $\eta_{\mathrm{heat}}=0.5$ and
%18\% for $\eta_{\mathrm{heat}}=0.25$. 
%than would be estimated if it was ignored.
Put another way, a significant rise in the mass transfer rate will
produce a 40\% decrease in $\dot{M}_{\mathrm{A}}(r_{\mathrm{disc}})$
because of heating by the accretion stream.  We expect standstill
luminosities about 40\% fainter than the peak of an outburst.

%------------------------------------------------------------
%============================================================
%------------------------------------------------------------
\section{Conclusion}

We have shown that inclusion of impact heating of an accretion disc by
the mass transfer stream from the secondary star has a significant
effect on the critical mass transfer rate required for stable
accretion. Accordingly, models attributing Z~Cam standstills to slight
increases of the mass transfer rate above this critical value do
predict that the standstills should occur at lower disc brightness
than the peak of the dwarf nova outbursts, as observed.

\section*{Acknowledgments}
A.R.K. thanks the UK Particle Physics and Astronomy Research Council
for a Senior Fellowship. R.S. and C.R. were supported by a PPARC
Rolling Grant for theoretical astrophysics to the Astronomy Group at
Leicester.  We thank H.~Spruit for valuable discussions.

\label{lastpage}
 
\end{document}